\documentclass[11pt,a4paper]{article}
\pdfoutput=1 
\usepackage{jinstpub}
\usepackage{units}
\usepackage{subfigure}

\usepackage{upgreek}
\graphicspath{{logos/}{figures/}}
\newcommand{\petalet}{petalet}
\newcommand{\petalets}{petalets}
\newcommand{\Petalet}{Petalet}
\newcommand{\Petalets}{Petalets}
\newcommand{\SplitRSfirst}{{\em split readout scheme}}
\newcommand{\SplitRS}{split readout scheme}
\newcommand{\CommonRSfirst}{{\em common readout scheme}}
\newcommand{\CommonRS}{common readout scheme}
\title{\boldmath Prototyping of \petalets\ for the Phase-II Upgrade of the silicon strip tracking detector of the ATLAS Experiment}
\author[a,j,1]{S.~Kuehn\note{Corresponding author.}}
\author[b]{V.~Ben\'{i}tez}
\author[b]{J.~Fern\'{a}ndez-Tejero}
\author[b]{C.~Fleta}
\author[b]{M.~Lozano}
\author[b]{M.~Ull\'{a}n}
\author[c]{H.~Lacker}
\author[c]{L.~Rehnisch}
\author[c]{D.~Sperlich}
\author[d]{D.~Ariza}
\author[d]{I.~Bloch}
\author[d]{S.~D\'{i}ez}
\author[d]{I.~Gregor}
\author[m,o]{J.~Keller}
\author[n,o]{K.~Lohwasser}
\author[d]{L.~Poley}
\author[d]{V.~Prahl}
\author[d]{N.~Zakharchuk}
\author[e]{M.~Hauser}
\author[e]{K.~Jakobs}
\author[e]{K.~Mahboubi}
\author[e]{R.~Mori}
\author[e]{U.~Parzefall}
\author[f]{J.~Bernab\'{e}u}
\author[f]{C.~Lacasta}
\author[f]{R.~Marco-Hernandez}
\author[f]{D.~Santoyo}
\author[f]{C.~Solaz Contell}
\author[f]{U.~Soldevila Serrano}
\author[i,k]{T.~Affolder}
\author[g]{A.~Greenall}
\author[h]{B.~Gallop}
\author[h]{P.W.~Phillips}
\author[l]{V.~Cindro}
\affiliation[a]{CERN, European Organization for Nuclear Research,\\ Geneve, Switzerland}
\affiliation[b]{Centro Nacional de Microelectr\'{o}nica (IMB-CNM, CSIC),\\ Campus UAB-Bellaterra, Barcelona, Spain}
\affiliation[c]{Institut f\"ur Physik, Humboldt-Universit\"at zu Berlin,\\ Newtonstra{\ss}e, Berlin, Germany}
\affiliation[d]{Deutsches Elektronen-Synchrotron,\\ Notkestra{\ss}e, Hamburg, Germany}
\affiliation[e]{Physikalisches Institut, Albert-Ludwigs-Universit\"at Freiburg,\\ Hermann-Herder-Stra{\ss}e, Freiburg, Germany}
\affiliation[f]{IFIC, CSIC-U. Valencia,\\ c/ Catedr\'{a}tico Jos\'{e} Beltr\'{a}n, Paterna, Spain}
\affiliation[g]{Department of Physics, University of Liverpool,\\ Cambridge Street, Liverpool, United Kingdom}
\affiliation[h]{Particle Physics Department, STFC Rutherford Appleton Laboratory,\\ Harwell Science and Innovation Campus, Didcot, United Kingdom}
\affiliation[i]{Santa Cruz Institute for Particle Physics (SCIPP),\\ University of California, Santa Cruz, USA}
\affiliation[j]{formerly at Physikalisches Institut, Albert-Ludwigs-Universit\"at Freiburg,\\ Hermann-Herder-Stra{\ss}e, Freiburg, Germany}
\affiliation[k]{formerly at Department of Physics, University of Liverpool,\\ Cambridge Street, Liverpool, United Kingdom}
\affiliation[l]{Jozef Stefan Institute Ljubljana,\\ Jamova cesta, Ljubljana, Slovenia}
\affiliation[m]{Carleton University, Department of Physics,\\ Colonel By Drive, Ottawa, Canada}
\affiliation[n]{University of Sheffield, Department of Physics and Astronomy,\\ Sheffield, United Kingdom}
\affiliation[o]{formerly at Deutsches Elektronen-Synchrotron,\\ Notkestra{\ss}e, Hamburg, Germany}
\emailAdd{susanne.kuehn@cern.ch}
 \abstract{
In the high luminosity era of the Large Hadron Collider, the HL-LHC, the instantaneous luminosity is expected to reach unprecedented values, resulting in about 200 proton-proton interactions in a typical bunch crossing. To cope with the resultant increase in occupancy, bandwidth and radiation damage, the ATLAS Inner Detector will be replaced by an all-silicon system, the Inner Tracker (ITk).
The ITk consists of a silicon pixel and a strip detector and exploits the concept of modularity.
Prototyping and testing of various strip detector components has been carried out. This paper presents the developments and results obtained with reduced-size structures equivalent to those foreseen to be used in the forward region of the silicon strip detector.  Referred to as \petalets, these structures are built around a composite sandwich with embedded cooling pipes and electrical tapes for routing the signals and power. Detector modules built using electronic flex boards and silicon strip sensors are glued on both the front and back side surfaces of the carbon structure. 
Details are given on the assembly, testing and evaluation of several \petalets. Measurement results of both mechanical and electrical quantities are shown. Moreover, an outlook is given for improved prototyping plans for large structures.
   }
\begin{document}

\maketitle
\flushbottom
\section{Introduction}
\label{sec:intro}
\begin{figure}[!htp]
\centering
        \includegraphics[width=0.9\textwidth]{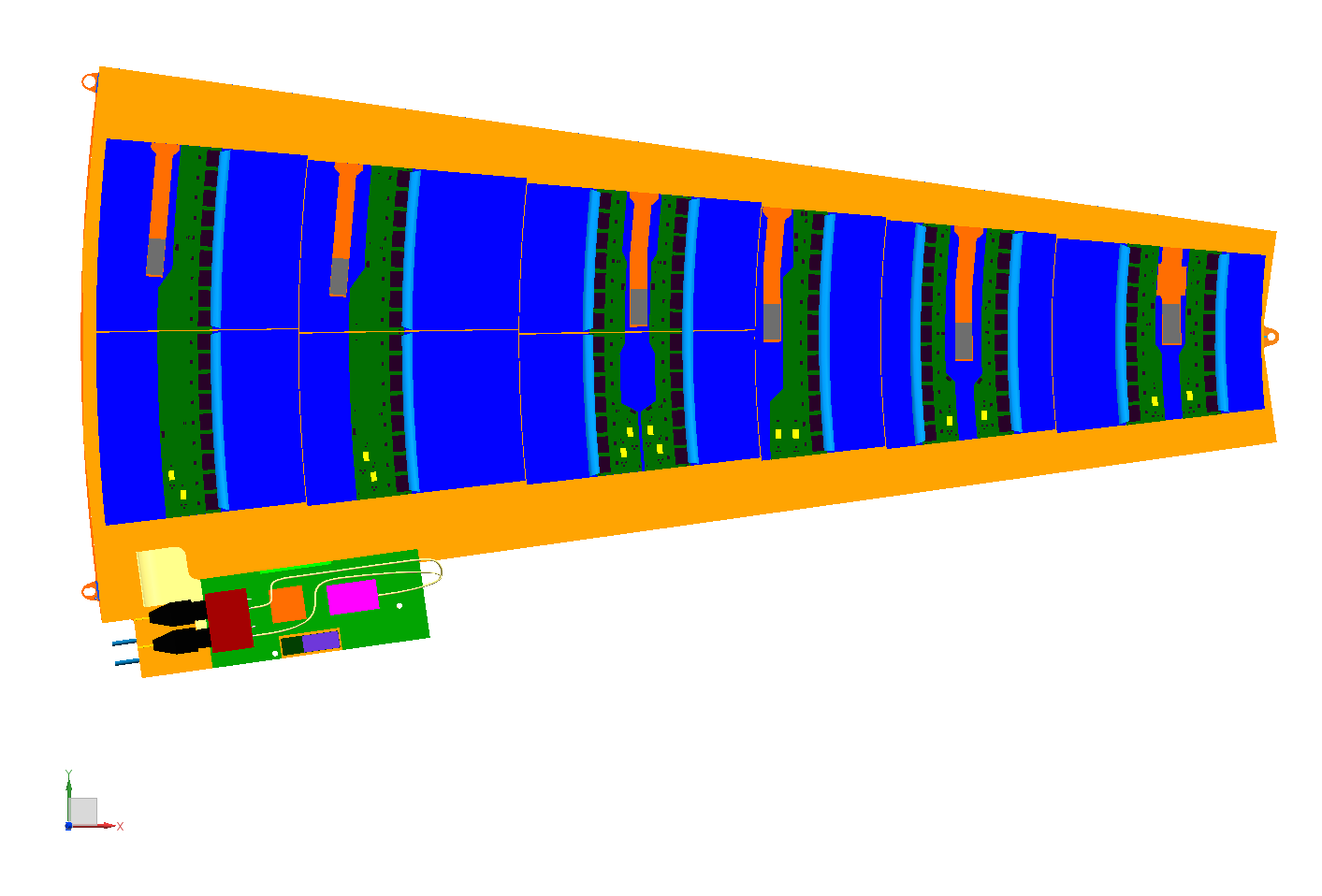}
                \caption{Schematic represenation of a petal consisting of a carbon-fibre core with cooling pipes inside. One side is visible with silicon-detector modules glued on top. The modules house the silicon sensor, readout-board and a powering board. Cooling pipe connectors and the readout board of a petal are visible at the side of the petal. The layout is for silicon sensors manufactured in 6" wafer technology and ASICs produced in a 130\,nm CMOS process.}
                \label{schematicPe}
\end{figure}

For the ATLAS experiment, a new, all-silicon tracker is foreseen at the High-Luminosity LHC (HL-LHC). 
It will consist of a pixel and strip detector~\cite{stripTDR} and is expected to be installed in 2024.
The strip tracking detector will feature five silicon sensor layers in the central and six in the forward region. An important concept of the upgrade tracker is modularity, allowing for faster final assembly and staged pre-assembly and pre-testing. 
For the central region so-called staves will be built and details of the prototyping can be found in Ref.~\cite{sergio}.
The basic building block used in the forward region is referred to as petal. They are double-sided objects in trapezoid-shape built from a carbon-fibre sandwich structure, the core and six ring sectors of silicon sensor modules glued onto both sides of a core. Figure~\ref{schematicPe} illustrates the geometry of a petal, including modules, silicon sensors (in blue) with readout electronics (in green and brown) glued on top. 
The end-cap modules are designed to cover the whole petal region in R-$\phi$ with two layers of castellated, slightly overlapping petals. The petal design is optimized for silicon sensors coming from 6" wafers. 
At higher R, in the regions in which a single silicon crystal coming from a 6" wafer cannot cover the whole area in $\phi$, sensors are split in two and placed next to each other.
In contrast, a single module provides coverage in the azimuthal direction for lower values of R.
In this region the modules employ the shortest strip sections to cope with the highest occupancy regions of the strip end-caps, thus requiring two readout boards per sensor. In addition, power boards (in orange in figure~\ref{schematic}) and the readout board of the modules and petal (in light green) are sketched. Titanium cooling pipes run inside the core and petal locators can be seen next to the outer ring of the petal, which compose the interface between the petals and the global end-cap structure. A multi-layered flexible circuit is assembled at the core surfaces (the so-called bus tape, in yellow), for the routing of power lines and electrical signals along the petal.

In total 384 petals with 6,912 modules will be produced. One disk of an end-cap will be assembled out of 32 petals.
A reduced-size version of petals, called \petalets\ have been developed to kick-off the prototyping of large structures with multiple silicon detector modules.
A detailed R\&D programme covering design, assembly and evaluation of these prototypes has been conducted. 
Two different readout architectures have been investigated to optimize the assembly and readout of full-size petals~\cite{sensorpaper,modulepaper,sven}.
The main results of this study, the assembly and evaluation of fully assembled \petalets\ are reviewed in this paper, including a comprehensive comparison between the two readout options investigated.

\section{Layout and components of \petalets}
\label{sec:petaletlayout}

In the following section the layout and the individual components of \petalets\ are described.

\subsection{\Petalet\ layout}
\label{subsec:Playout}

\begin{figure}[!htp]
\centering
        \includegraphics[width=0.7\textwidth]{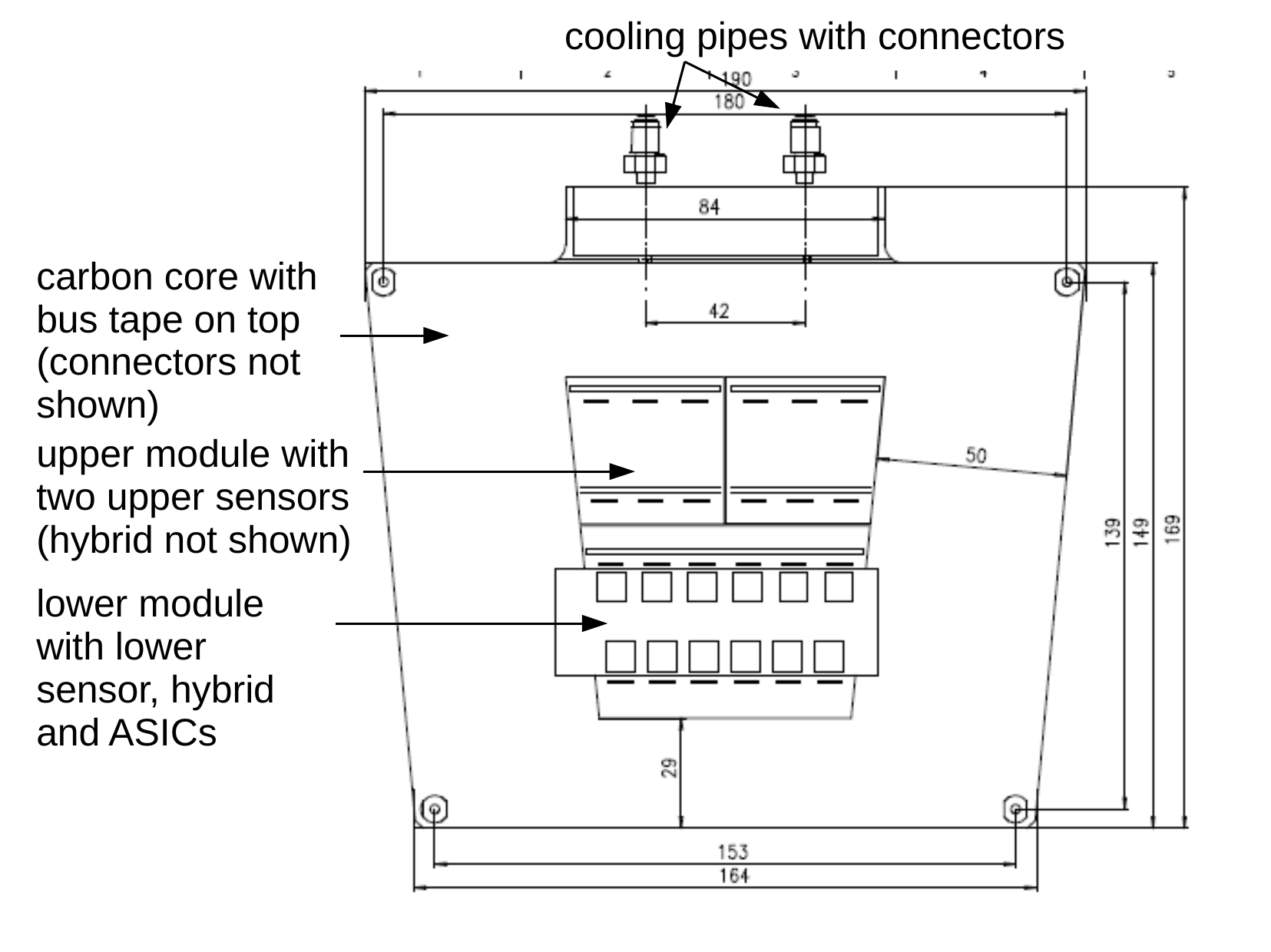}
                \caption{Schematic drawing of a \petalet\ consisting of a carbon-fibre core with cooling pipes inside. One side is visible with two upper sensors and a lower module glued on top. For the upper module the hybrid is not shown but lines indicate the placement and wire-bond area. For the lower module the hybrid is drawn on top. Dimensions are shown in millimeters. Taken from Ref.~\cite{modulepaper}.}
                \label{schematic}
\end{figure}

The \petalets\ are smaller-size prototypes and consist of similar components to the ones which are used in petals. Their overall size is 15\,x\,18\,cm$^{2}$ and they are about 6\,mm thick.  
Instead of nine sensors like on a petal, three silicon sensors are glued on each side of a \petalet\ core with their respective readout electronics, consituting fully functional silicon microstrip modules. A schematic sketch with core dimensions in millimeters is given in figure~\ref{schematic}.
It sketches the top surface of the carbon-fibre core attached with a lower, big sensor and two smaller top sensors having flex printed-circuit-boards, called hybrids, glued on top. The hybrids have readout chips designed in 250\,nm CMOS process glued on top which were the version available at the time of the protoyping phase. The ASICs have 128 analogue channels with binary output allowing to process signals from the silicon strip sensors in a binary read-out architecture. The readout channels and silicon strips are connected by wire-bonds.
Titanium cooling pipes are placed inside the carbon-fibre core and a thin (150\,$\mu$m) multi-layer flexible circuit tape for electrical routing is glued on top of the core surfaces. The connectors of the pipes can be seen sticking out the core. Additional boards required for the electrical readout of the modules are glued on the tape.
With these double-sided support structures the routing between front and back side as well as the handling and assembly can be prototyped.

\begin{figure}[!htbp]
\begin{center}  
        \includegraphics[width=0.45\textwidth]{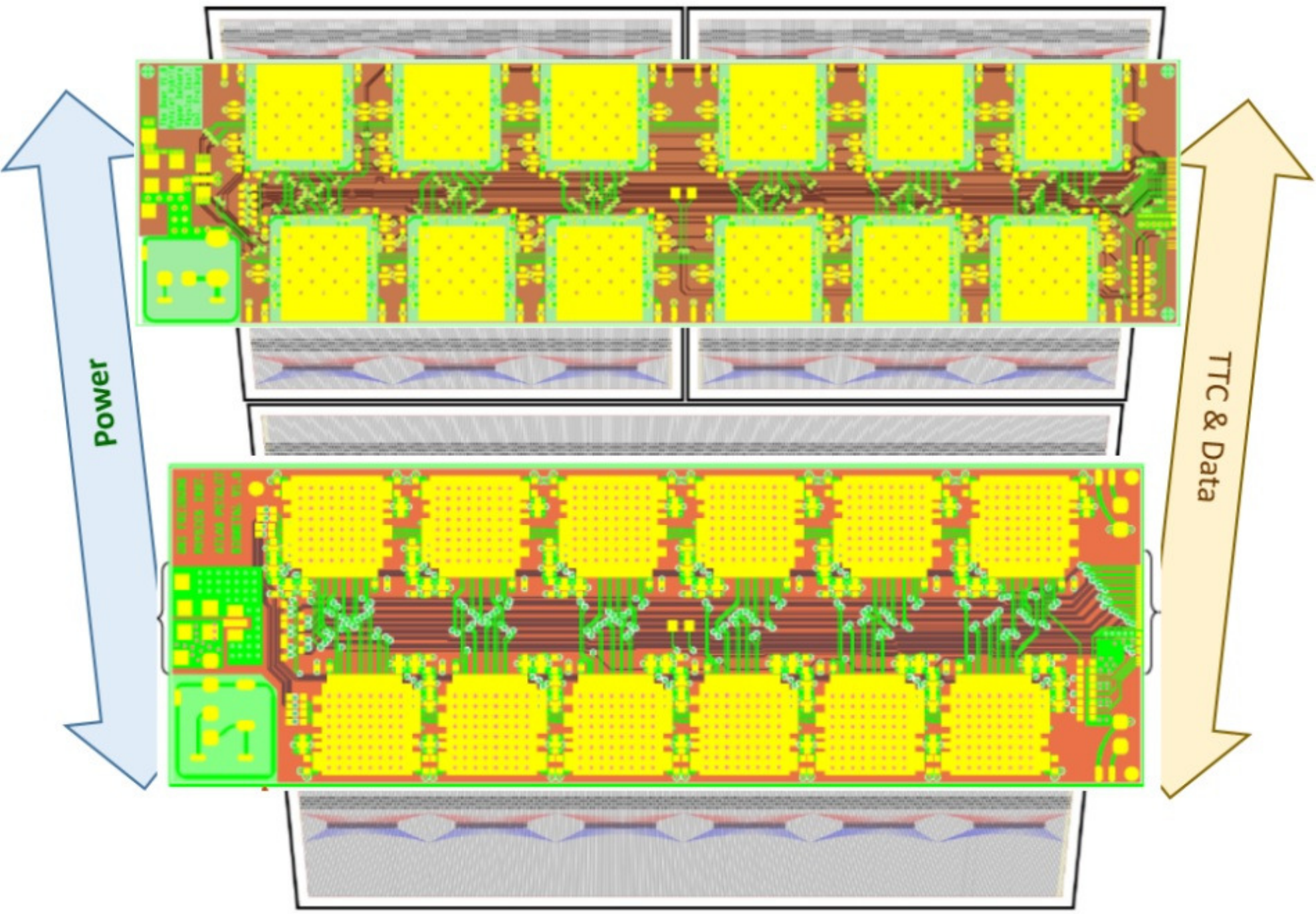}
       \caption{Sketch of one side of the \SplitRS\ layout. Two modules, each consisting of one hybrid mounted on one (lower module) or two (upper module) sensors, respectively, are placed and read out on one bus tape on each side of a \petalet. Each hybrid hosts 12 chips. The power is routed on the left-hand side whereas data, clock, and timing, trigger and control (TTC) signals are routed on the right-hand side of the hybrids, modules, respectively. Taken from Ref.~\cite{modulepaper}.}
        \label{fig:BearHybridPetaletSketchFront}
\end{center}    
\end{figure}

\begin{figure}[!htb]
        \begin{center}
        \includegraphics[angle=270,width=0.45\textwidth]{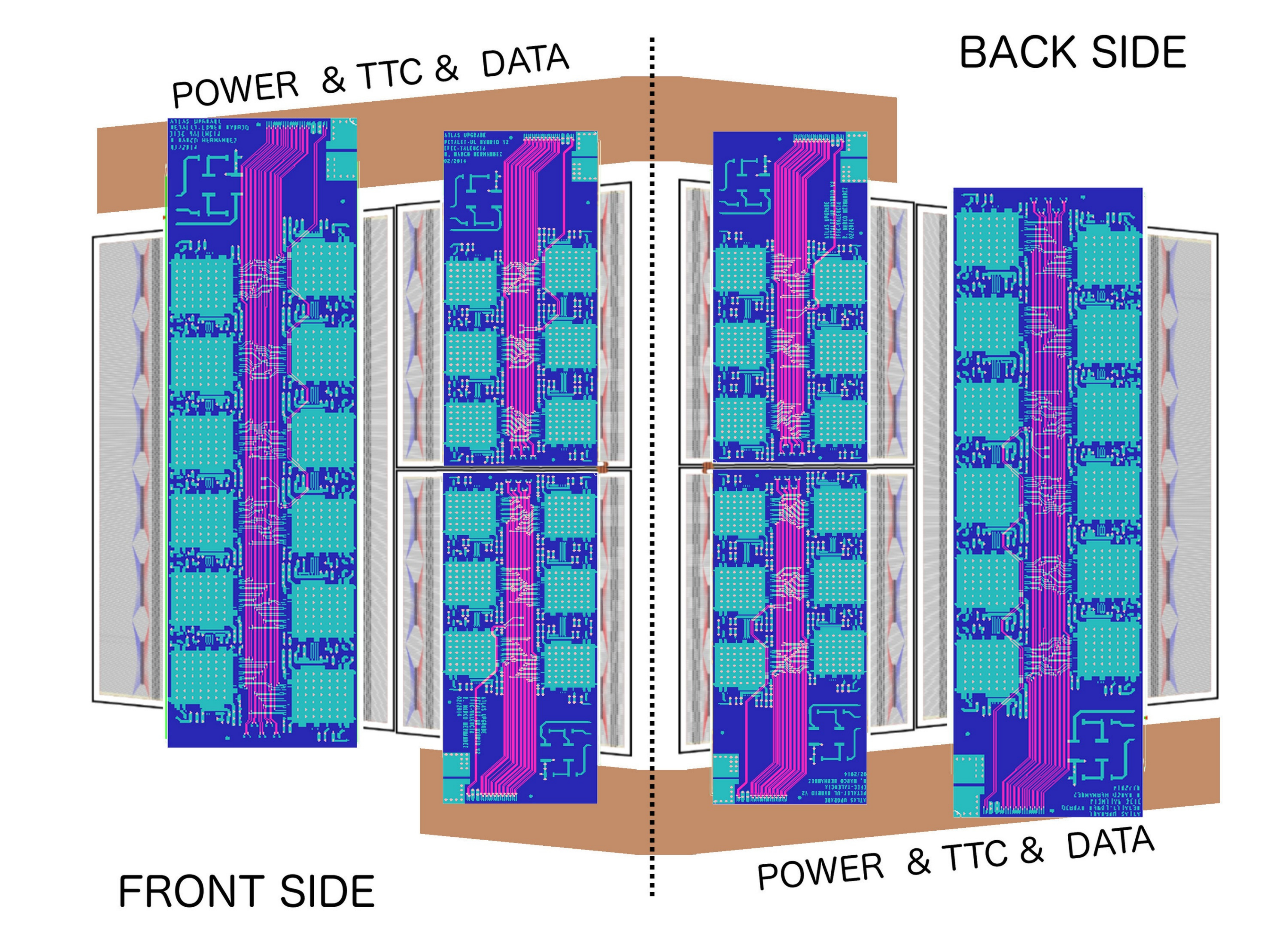}
                \caption{Sketch of the two sides of the \CommonRS\ layout. Six hybrids are used to read out six sensors. They host either six or twelve chips. A single piece of bus tape covers both the front and back side of the \petalet, providing power and data lines for both sides. The bus on the left-hand side connects one frontside and two back side modules, while the bus on the right-hand side connects one back side and two front side modules. Taken from Ref.~\cite{modulepaper}.}
                \label{LFhscheme}
\end{center}
\end{figure}

During the prototyping of the \petalets, two different readout architectures have been under investigation. They are compared aiming for low complexity and high performance. 
One readout scheme deploys connections and routing of data, clock, and trigger, timing and control signals at the right-hand side of the module, while powering is routed on the left-hand side. This scheme is in the following called \SplitRSfirst. A schematic sketch, indicating the routing, can be seen in figure~\ref{fig:BearHybridPetaletSketchFront}. The second scheme routes data, trigger and control, and powering to one side of the module. This one is called \CommonRSfirst\ and sketched in figure~\ref{LFhscheme}. A single piece of bus tape covers both the front and back side of the \petalet, providing power and data lines for both sides. The bus on the left-hand side connects one front side and two back side modules, the one on the right-hand side one back side and two front side modules.
Six hybrids are used to read out six sensors. 
Specific tooling has been developed for all dedicated construction and assembly steps for both layouts.

\subsection{Modules}
\label{subsec:modules}
Silicon sensor modules have been developed within the \petalet\ project. The modules consist of a flex printed circuit board with ASICs, the hybrids glued on top of one or two silicon sensors. A detailed description is given in Ref.~\cite{modulepaper,kambiz}.
The ASICs are produced in 250\,nm CMOS process and specifications are given in Ref.~\cite{kaplon,abcn}. In the future, they will be produced in a 130\,nm CMOS process, with 256 readout channels per ASIC, which leads to a reduced number of ASICs per silicon module for petals as opposed to the \petalets~\cite{abc130}.
 The sensors are made of p-type bulk material with p-stop insulation~\cite{CNM,sensorpaper}.
Three sensor geometries have been designed:  ``lower big'', ``upper left'', and ``upper right''. In addition, two sensor types are available, a set with embedded pitch adapters to reduce the wire-bonding angle and a set with standard pitch adapters on each strip. The second metal layer of the embedded pad sensors leads to increased noise values depending on the length and geometry of the second metal layer traces~\cite{embedded_upgrade, hiroemb}.
These effect has to be taken into account when comparing the performance of \petalets\ with modules built either from standard sensors or from embedded ones. 

Both readout schemes require different hybrid layouts. They vary in both the direction of data and power readout as described before and the amount of chips per hybrid, either six or twelve. 
In addition, the upper modules of the \SplitRS\ have one hybrid glued onto two silicon sensors. For the \CommonRS, individual hybrids are glued on the upper left and upper right sensor, resulting in total in three modules per \petalet\ side.
Details on the hybrids for the forward region are given in Ref.~\cite{modulepaper,kambiz}. In these papers also the assembly and testing of modules in test frames is described. 
In total about 40 modules have been built and thoroughly tested. The test results show a good performance and low average input noise values of 380$\pm$20\,ENC for hybrids and 520-750\,ENC for modules depending on the sensor quality.

\subsection{\Petalet\ cores}
\label{subsec:core}
\begin{figure}[h!]
  \centering
  \includegraphics[width=0.5\textwidth]{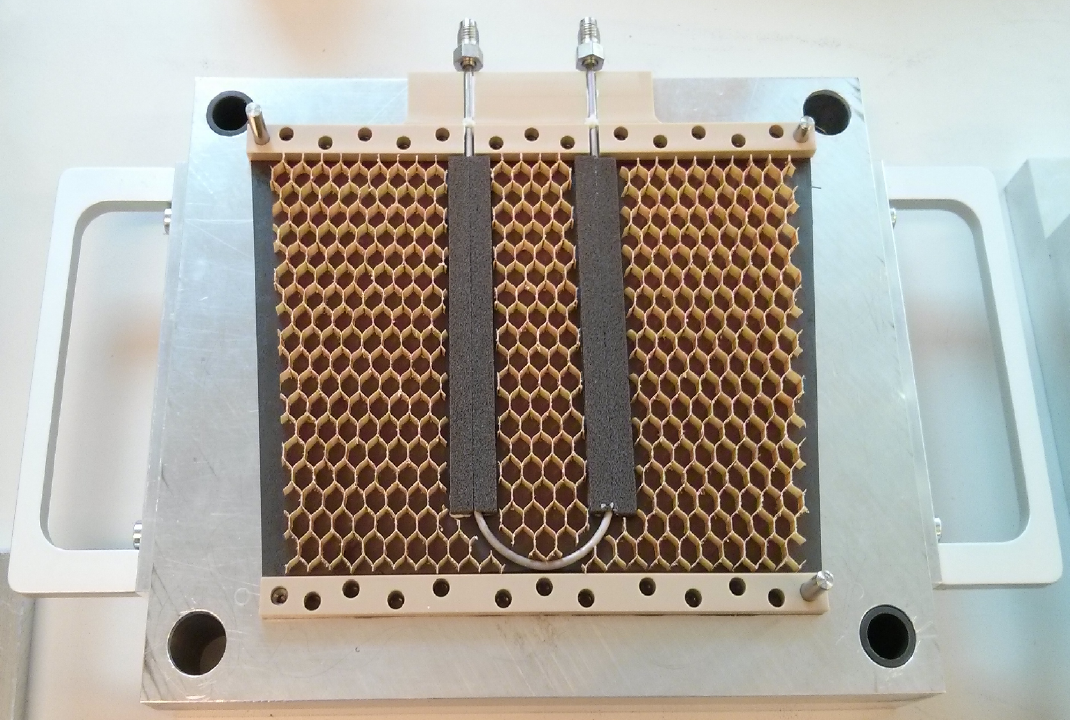} 
  \caption{Petalet\ core on a mounting jig without the top CFRP skin.}
  \label{coresA}
\end{figure}

\begin{figure}[h!]
  \centering
  \includegraphics[width=0.5\textwidth]{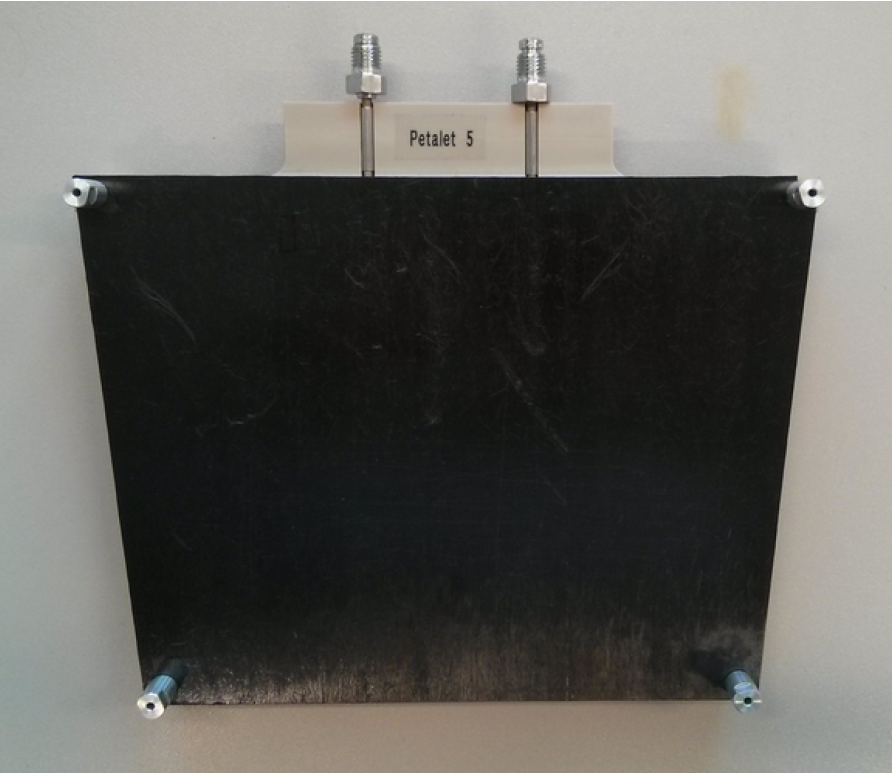} 
  \caption{Completed core structure.}
  \label{coresB}
\end{figure}

The \petalet\ core structure mimics the envisioned design for the full-length support structures, the petals. It consists of a flat, lightweight carbon-fiber based structure with embedded titanium cooling pipes. The choices of materials and adhesives aimed for the best compromise between thermal and mechanical performances, also after irradiation, and a minimum material budget. The ``skins'' of the structure are constituted of a 3-layers layup of carbon fiber reinforced polymer (CFRP), co-cured together with a certain fiber orientation among them (0-90-0 degrees, from top to bottom). This fiber orientation was found to provide the optimal thermal and mechanical properties of the petal structures. The titanium cooling pipe is bent to an U-shape and has an outer diameter of 2.275\,mm, with 140\,$\mu$m wall thickness. Stainless steel fittings are brazed into the pipe inlet and outlet. Inside the core structure, the pipe is surrounded by thermally conductive, carbon fiber-based foam of very low density (0.23\,g/cm$^{3}$), from Allcomp Inc.~\cite{allcomp}. The pipe-foam structure is glued on the back side of the CFRP skins with Boron Nitride-loaded Hysol EA9396 epoxy from Henkel Inc.~\cite{henkel}. The rest of the structural material consists of Kevlar N636 Para-Aramid fiber honeycomb from DuPont Inc.~\cite{dupont}. The honeycomb is also glued down to the top and bottom CFRP skins with EA9363 epoxy. The core structure is closed at its sides with closeout elements of different shapes, machined in thermoplastic polymer (PEEK) and glued to the skins with EA9363. An electrical break, required to isolate electrically the petals from noise sources from the outside world, will be included in petal pipes but is not yet in the \petalet\ cores. Custom mechanical tools were developed for the precise assembly of the \petalet\ cores. 
Figure~\ref{coresA} shows a photo of the internal structure and figure~\ref{coresB} a fully completed core with the top face-sheet attached. The total thickness of the \petalet\ cores, including the bus tape (described in section~\ref{subsec:bustape}) on top, is 5.4\,mm. Its mass is about 77\,g. Nine \petalet\ cores have been manufactured for the \petalet\ project.

\subsection{Bus tape}
\label{subsec:bustape}
\begin{figure}[!htp]
\centering
        \includegraphics[width=0.6\textwidth]{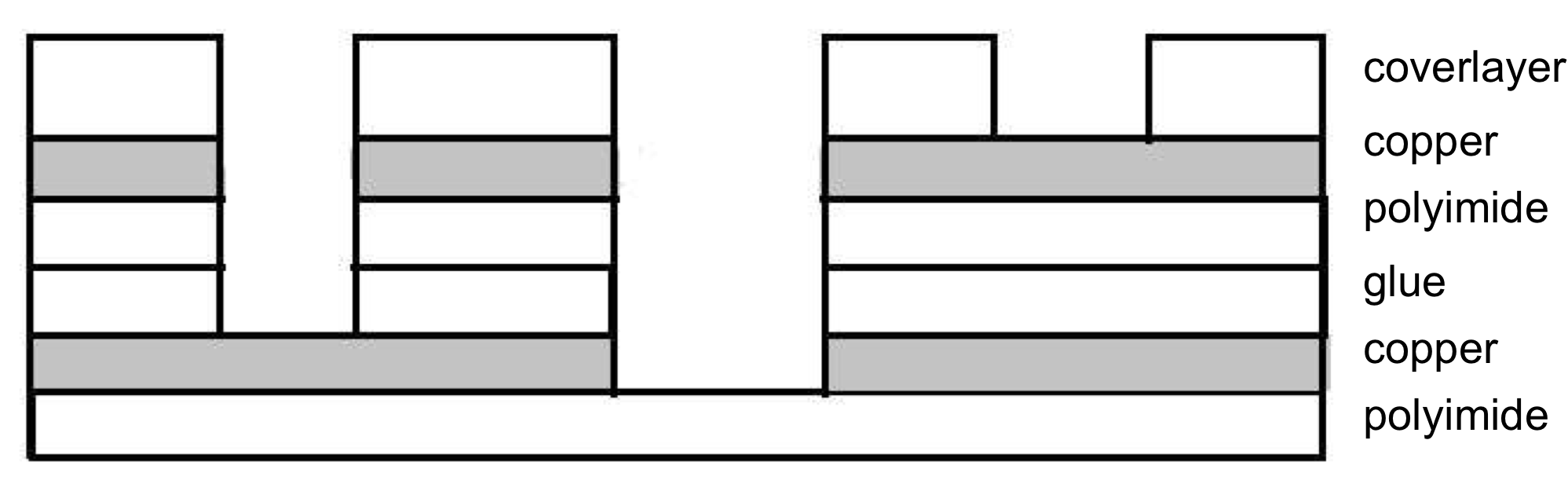}   
                \caption{Schematic sketch of the stack-up of the bus tapes.} 
                \label{sketchtape}
\end{figure}

\begin{figure}[!htp]
\centering
        \includegraphics[width=0.7\textwidth]{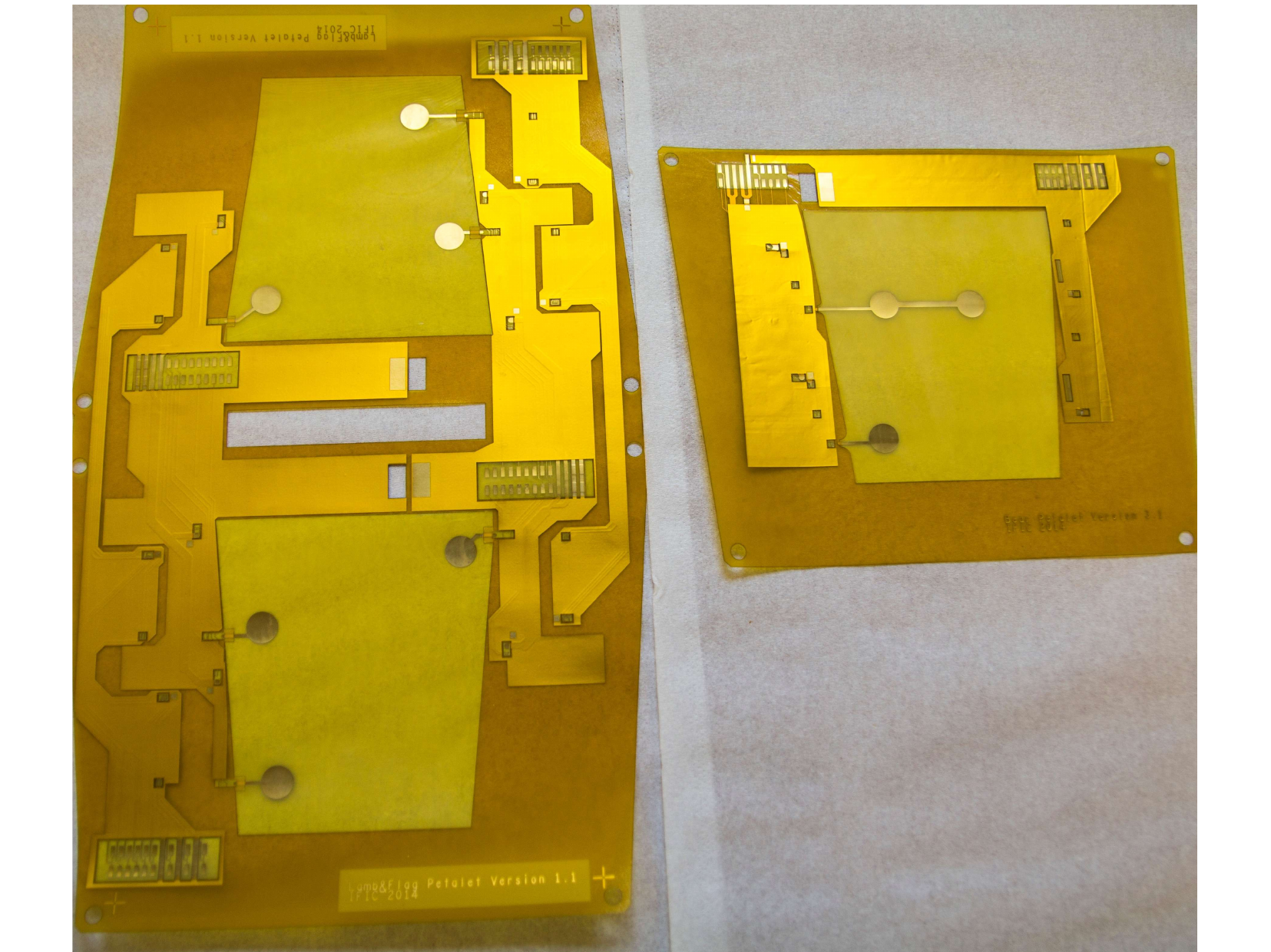}
                \caption{Photographs of bus tapes for both readout schemes. For the \CommonRS\ on the left, it covers both front and back side of the core and is bent over the upper part of the core. On the right, the bus tape for the \SplitRS\ which deploys one for each side of the \petalet.}
                \label{tapebear}
\end{figure}

The bus tape is a printed circuit on a flexible substrate. Its main purpose is to route power, command and data lines between the front-end electronics and the outer data acquisition system. The bus tape is aimed for low material budget while giving good thermal path 
between the hottest elements (front-end electronics and sensors when irradiated) and the cooling pipes that run inside the carbon fibre core. Bus tapes are designed to be manufactured in thin polyimide at ELGOLINE d.o.o.~\cite{elgo}. Furthermore, bus tape designs for the \petalet\ minimise the number of polyimide layers to one in the region under the sensors and a maximum of two metal layers (copper traces under a metal shield) is used.
For simplicity and ease of manufacturing, vias are avoided. Connection to the bottom metal layer is done with openings and wirebonding.
A schematic sketch of the stack-up of the bus tapes can be seen in figure~\ref{sketchtape}.
In the first prototypes, aluminium was used as a conductor in the ground-shield layer to minimize the scattering material. However it was abandoned due to the mismatch of the coefficient of thermal expansion with the copper layers, leading to deformations in the tape during manufacture. 12\,$\mu$m copper - 25\,$\mu$m polyimide laminate, 25\,$\mu$m glue joining layers are used in the later versions for both layers. Openings in polyimide and glue are done with a laser cutter. Wire-bondable electrodeposited gold (ENIG) is covering the exposed areas of copper traces. 12.5\,$\mu$m polyimide with 25\,$\mu$m glue coverlayer protects the upper surface of the tape. 
Power lines to the front-end electronics and sensors consist of two power supply lines (about 10\,V), two control lines for power supply of the DC-DC converter (more in Section~\ref{subsec:setup}) and one pair of high voltage line (HV) and its return (HVret) per sensor. Widths of power lines and separation of HV lines are designed following IPC-2221 standards~\cite{IPC}.
Command and control lines require three differential pairs (COM, BCO, L1R). A DATA differential pair and two lines to read a temperature sensor connect to each hybrid.
As the carbon-fibre core has low resistivity, an opening is designed in the bus tape in order to ground the core. 
The bus tape plays an important role in the evaluation of the two readout options for the forward region, not only electrically but also mechanically. Thus, two bus tape designs are implemented and photographs of both are shown in figure~\ref{tapebear}.
On the left photograph, the one for the \CommonRS\ is displayed. In this option, each sensor has its own hybrid. 
 The design covers both \petalet\ sides and it is folded around the upper part of a core. 
The photograph on the right presents the bus tape for the \SplitRS. 
Two equal bus tapes, one on each side of the \petalet\ core, are needed for this readout scheme. 
Soldering pads for connectors are visible on both tapes.

\subsection{Electrical boards}
\label{subsec:setup}

For powering and data readout, the \petalets\ are connected to the so-called End-of-\petalet\ boards. Two types of these boards have been developed for the two readout architectures. The boards filter and multiply data to the HSIO readout board~\cite{hsio} which is connected to a computer with data aquisition software. 
In addition, two more types of boards are glued onto the \petalet. One buffer control chip (BCC) per module, mounted on a printed circuit board, is used as an interface board to the ASICs on the hybrid and fosters signal distribution. By this the hybrids are addressed and the number of required signal lines is reduced. It also handles trigger, timing and command signals. The second ones, low noise and low mass DC-DC converters, allow powering of the hybrids. The converters provide 2.5\,V to the front-end chips from the 10\,V supply voltage. Each converter uses a toroidal coil inductor, which needs to be shielded to reduce the emitted electromagnetic noise. DC-DC powering is the baseline for strip modules in the upgrade tracker.
Both boards can be seen assembled on a \petalet\ in figure~\ref{petalassem}.

\section{Assembly of \petalets}
\label{sec:assembly}

In the following section, the assembly of \petalets\ is described.

\subsection{Assembly and testing of bus tape and core}
\label{subsec:assemcoretape}
\begin{figure}[h!]
  \centering
  \includegraphics[width=0.7\textwidth]{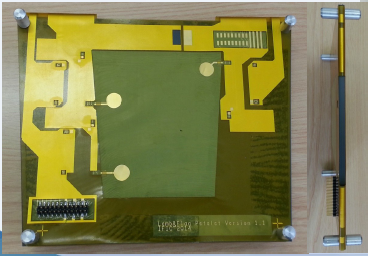}
  \caption{Test assembly of a dummy core with bus tape of \CommonRS\ glued and bent over the upper part of the core. On the right-hand side a top view of the bent area is visible.}  
  \label{coresC}
\end{figure}

\begin{figure}[!htp]
\centering
        \includegraphics[width=0.7\textwidth]{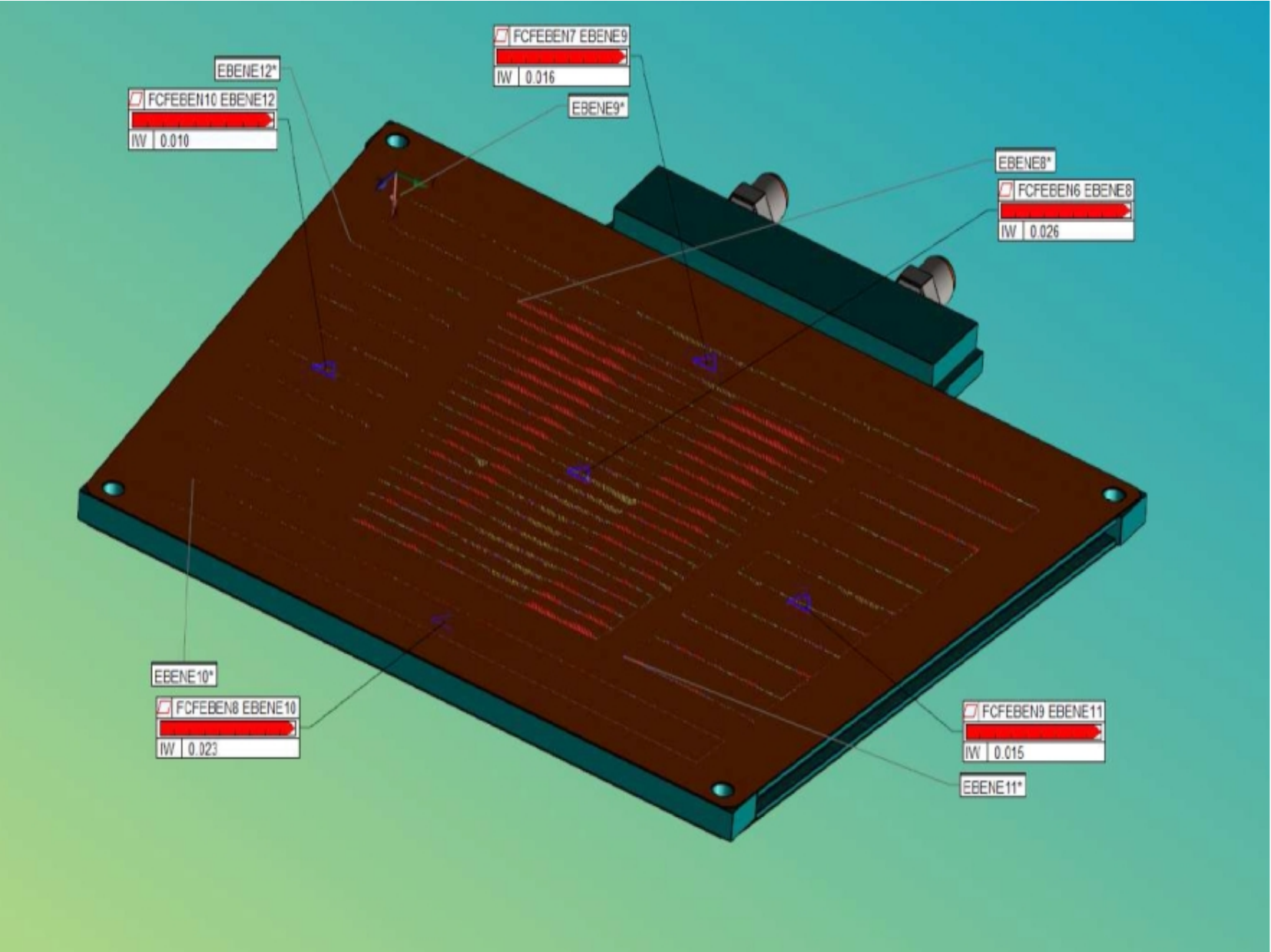}
                \caption{Results of a planarity measurement of a representative \petalet\ core. Surface is divided in five different sectors, due to the different thicknesses of the bus tape: silicon area (center), top area, lateral areas, and bottom area. Average flatness tolerance is 37$\pm$12\,$\mu$m.}
                \label{planarity}
\end{figure}

Once the cores are finished, the bus tape with 150\,$\mu$m thickness is glued onto the top side of both CFRP skins with a thin layer of EA9363 epoxy. Figure~\ref{coresC} shows a core with the bus tape of the \CommonRS.
On the right the bent tape area is displayed.
In the baseline petal layout, the bus tapes are co-cured together with the CFRP skins and no additional glue layer or step is required. 
The planarity of cores including bus tapes is measured after curing and figure~\ref{planarity} shows the result of a planarity measurement of a representative core structure performed with a contact probe mounted on a coordinate measuring machine (CMM). The measurements were divided in sectors, since the bus tapes have different thicknesses over the core surface. In the region below the silicon sensors, the average measured thickness is 5.39$\pm$0.05\,mm. The finished cores exhibit a flatness tolerance equal to 37$\pm$12\,$\mu$m. 
In total six cores were successfully assembled with bus tapes. In a next step, connectors are soldered on bus tapes and fully functional modules are mounted after cleaning the surface with alcohool.

\subsection{Loading of modules onto cores}
\label{subsec:assemcoretape}

\begin{figure}[h!]
  \centering
    \includegraphics[width=0.6\textwidth]{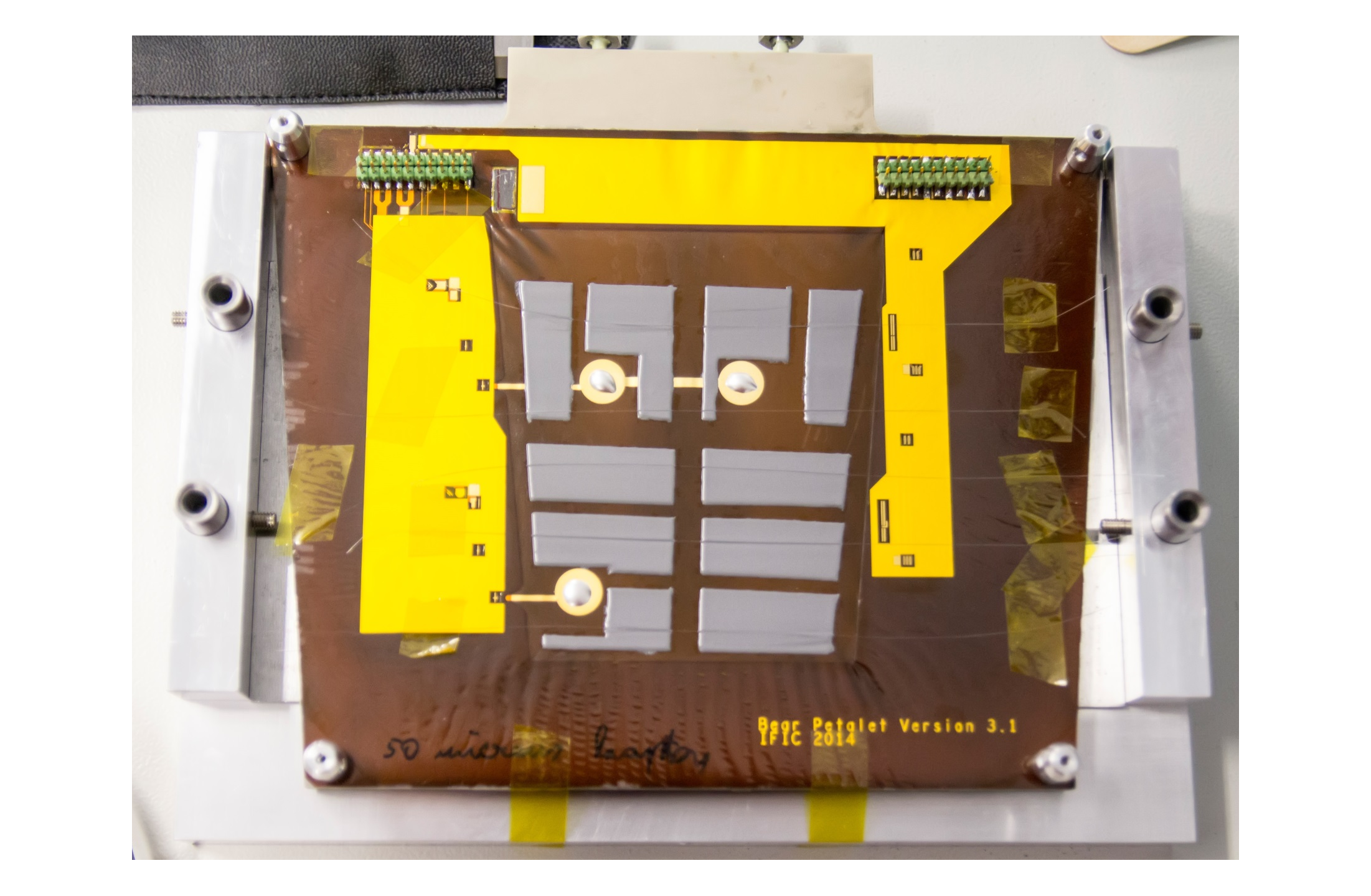}
  \caption{Glue pattern on core surface with a \SplitRS\ bus tape before module assembly.}
  \label{assemblyA}
\end{figure}

\begin{figure}[h!]
    \centering
    \includegraphics[width=0.7\textwidth]{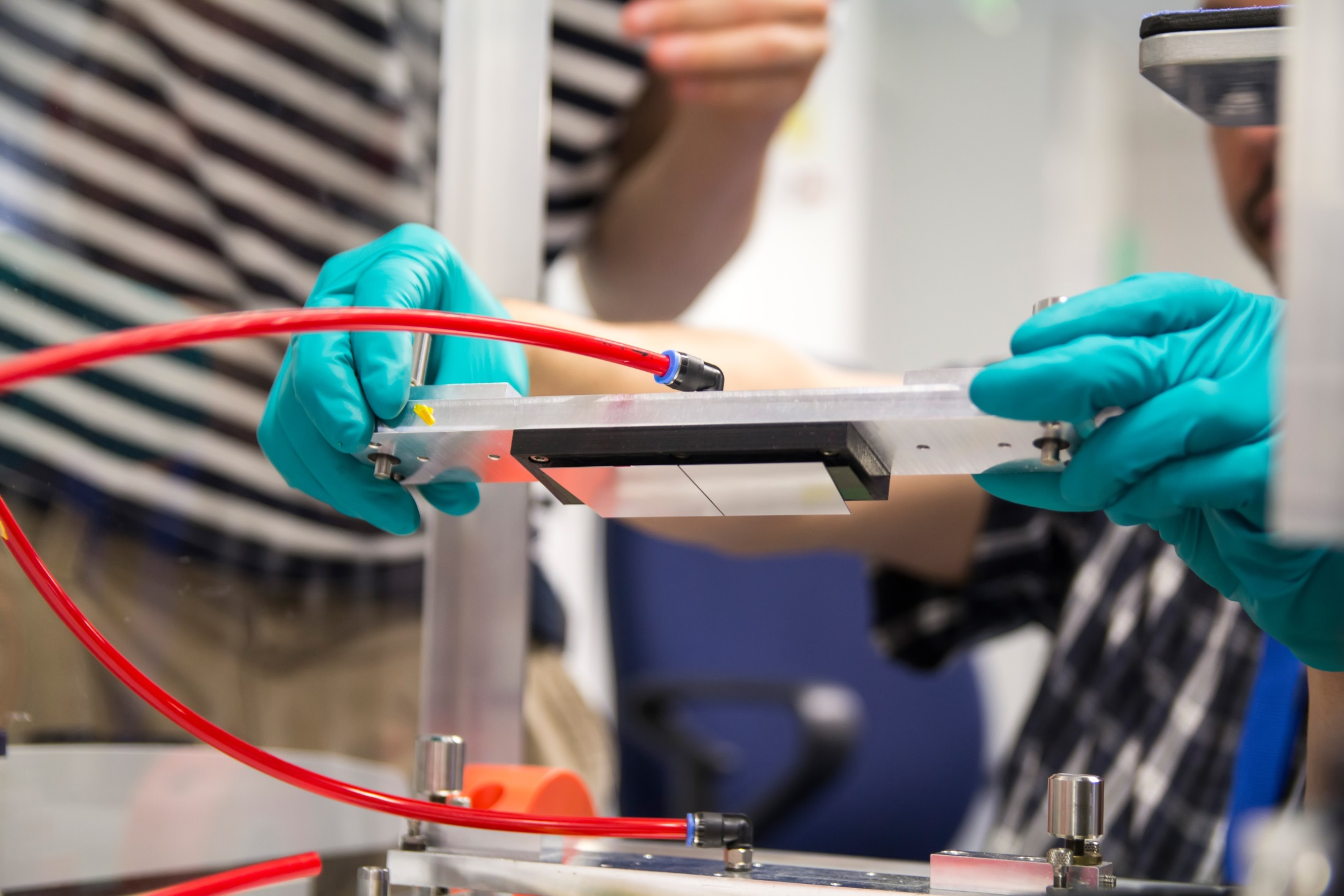}
    \caption{Module pick-up with custom pick-up tools.}
                \label{petalassemB}
\end{figure}

\begin{figure}[h!]
    \centering
    \includegraphics[width=0.7\textwidth]{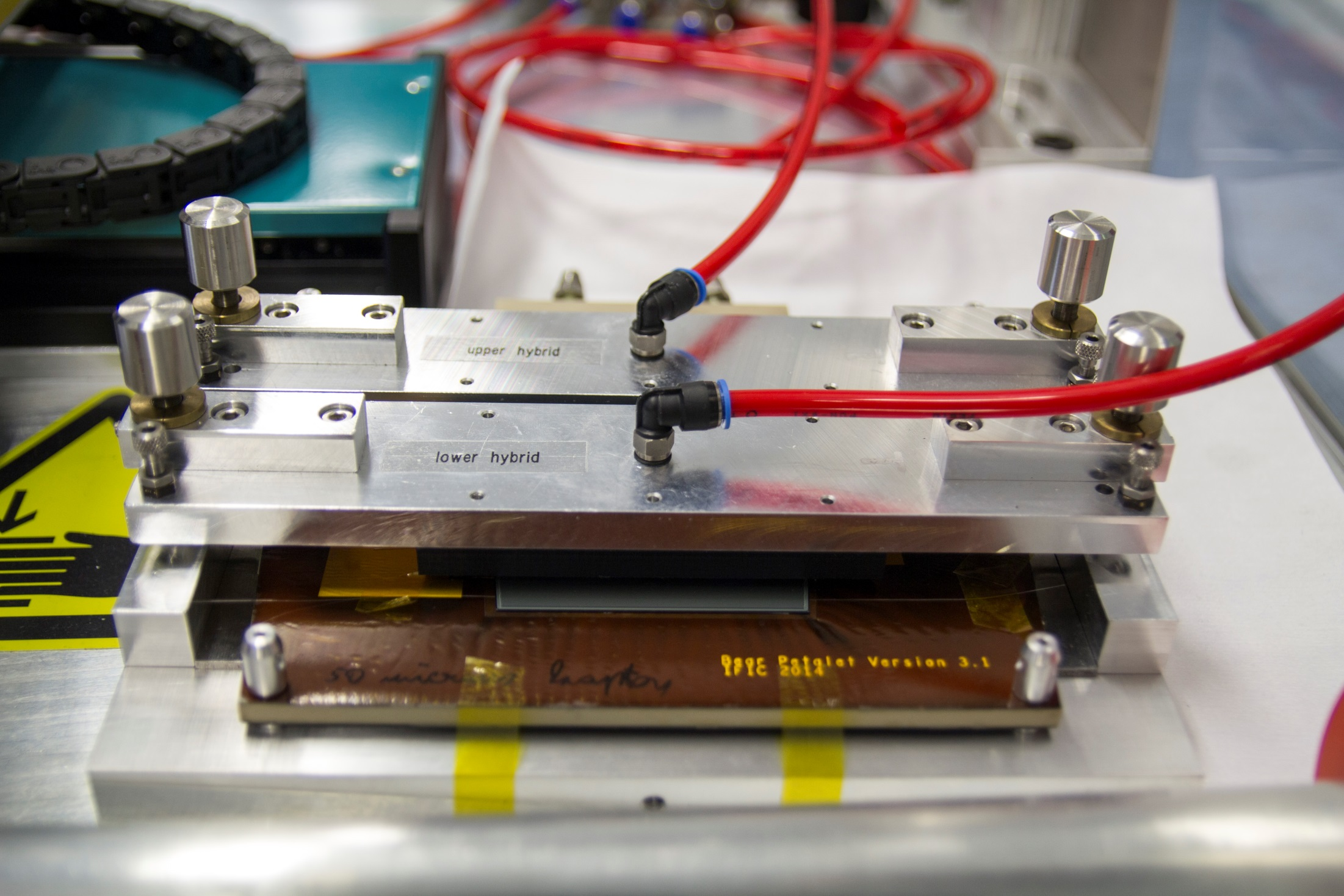}
    \caption{Module placement on core with custom pick-up tools.}
                \label{petalassemC}
\end{figure}

A custom set of tools was developed for the precise placement of the \petalet\ modules onto the \petalet\ cores. A set of vacuum plates and vacuum-holding pick-up tools is used to pick up the silicon modules once they have been disconnected from their test frames. The modules are located on a vacuum plate integrated with an X-Y-stage driven by micrometer screws. Pick-up occurs on the ASICs surface, preserving enough clearance to the wire-bonds. Fiducials located on the silicon sensors allow for precise pick-up of the modules onto the \petalet\ core. The \petalet\ core is placed on a custom precision frame, to which the pick-up tools holding the modules are then mechanically locked. The modules are glued down onto the \petalet\ cores with SE4445 thermally conductive and electrically insulating silicone gel from Dow Corning Inc.~\cite{dowcorning}. The high voltage (HV) contact of the sensor backplane to the traces on the bus tapes is achieved with electrically conductive TRA-DUCT 2902 epoxy from Henkel Inc.~\cite{henkel}. The glue is dispensed on the core surface by means of a custom glue stencil template. The stencil is designed to maximize the SE4445 glue spread and hence the thermal contact between the \petalet\ core and the strips modules. The glue pattern on the core surface can be seen in figure~\ref{assemblyA}. The thickness of the glue is determined by 150\,$\mu$m diameter nylon wires and with adjusted micrometer screws on the pickup tools.  Once the modules are glued, the DC-DC converters and BCC boards are also assembled onto the \petalet\ core with thermally conductive glue. Figures~\ref{assemblyA},~\ref{petalassemB} and~\ref{petalassemC} show different stages of a \petalet\ undergoing the gluing of the silicon modules with the custom mechanical tools.
Figure~\ref{petalassemB} shows the module pick-up and figure~\ref{petalassemC} the final placement onto the core with custom pick-up tools. Both an upper and a lower module are placed on the core for curing. The clearances between sensors are 500\,$\mu$m.
\begin{figure}[h!]
    \centering
    \includegraphics[width=0.7\textwidth]{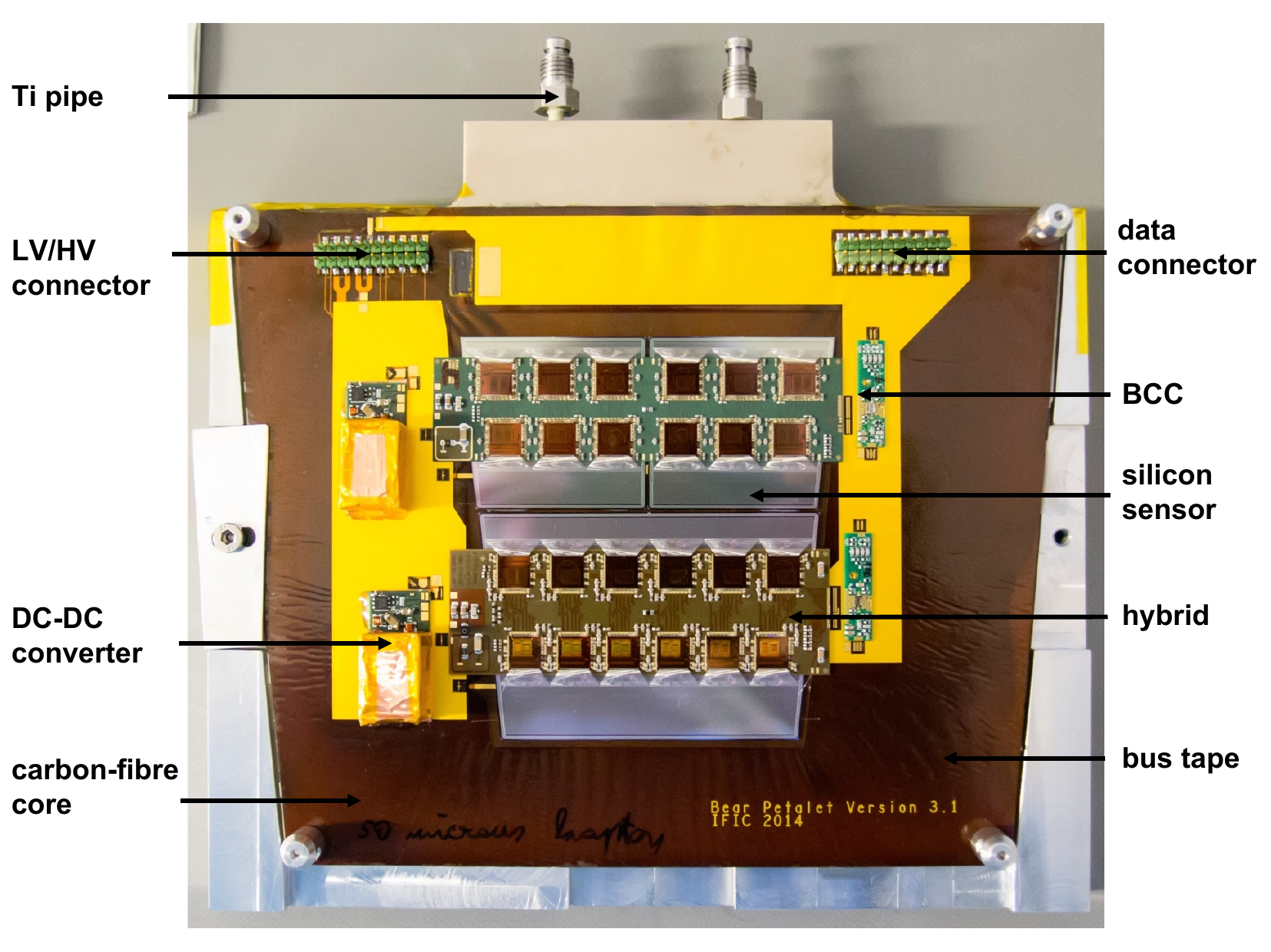}
    \caption{Completed \petalet\ including DC-DC converter and BCC boards. Two upper and one lower module of the \SplitRS\ are glued onto the bus tape.}
                \label{petalassem}
\end{figure}
Figure~\ref{petalassem} shows one side of a fully assembled \petalet\ in the \SplitRS.
Before electrical testing of the \petalets, hybrid and bus tape pads need to be connected via wire-bonds. 
All cores with bus tapes are assembled to five \petalets\ in the \SplitRS\ and one in the \CommonRS.

\section{Testing of \petalets}
\label{sec:testing}

A variety of tests were conducted with the \petalets.
Both \petalets\ with standard and with embedded modules and \petalets\ of both readout schemes are evaluated as presented in the following sections.
 The results are compared between different types of \petalets. Moreover, electrical test results of modules are compared before and after gluing on the core as well as measurements were performed at different operating temperatures. 

\subsection{Test procedure for electrical tests}
\label{subsec:testproc}
The \petalet\ under test is connected using custom-designed adapter boards and cables to low voltage and high voltage power supplies. The connection of data lines is made with a flat-ribbon cable to an HSIO readout board~\cite{hsio} which provides the interface to the computer. 
For the \SplitRS\ \petalets\ both sides are connected individually to the adpater boards, while for the \CommonRS\ \petalet\ the bus tape connects both sides and they are connected to one adapater board and cabling.

A first test is the verfication of the connectivity of all ASICs of the modules by requesting their addresses.   
Afterwards, the important quantities like noise behavior and signal readout of the modules are tested with the HSIO test board and specific data acquisition software. At the beginning, the delays are correctly set. Then every channel of the ASICs is evaluated using threshold scans in the binary readout architecture. In a threshold scan a number of hits at a fixed injected charge versus the discriminator threshold voltage is recorded. By varying the threshold and repeating (in the measurements 200 times), the occupancy curve, the number of hits above threshold versus threshold value can be obtained. Due to noise contributions this curve is widened by a Gaussian distribution and resulting in a curve called s-curve. The output noise is the sigma extracted from the width of the s-curve. The 50\% occupancy point of the s-curve for various values of injected charges leads to a curve called response curve distribution. Its derivative corresponds to the gain of the channel. By dividing output noise and gain, the input noise at the discriminator in ENC (equivalent noise charge) can be calculated. To reduce the spread of the gain in the individual channels, a trim circuitry in the chips can be used. 
The mean of the input noise per chip is the most important quantity in following comparisons of modules before and after assembly on petalets.
It is obtained from a Gaussian fit of the input noise values from each individual channel of the ASIC. Its uncertainty is the standard deviation of the fit over the 128 channels~\cite{modulepaper}.

\subsection{Testing of \SplitRS\ \petalets}
\label{subsec:testbear}

Five \SplitRS\ \petalets\ were evaluated. Both the input noise and its stability was measured under different conditions and the thermal behavior of a \SplitRS\ \petalet\ was investigated.

\subsubsection{Electrical tests}
\begin{figure}[h!]
    \centering
    \includegraphics[width=0.7\textwidth]{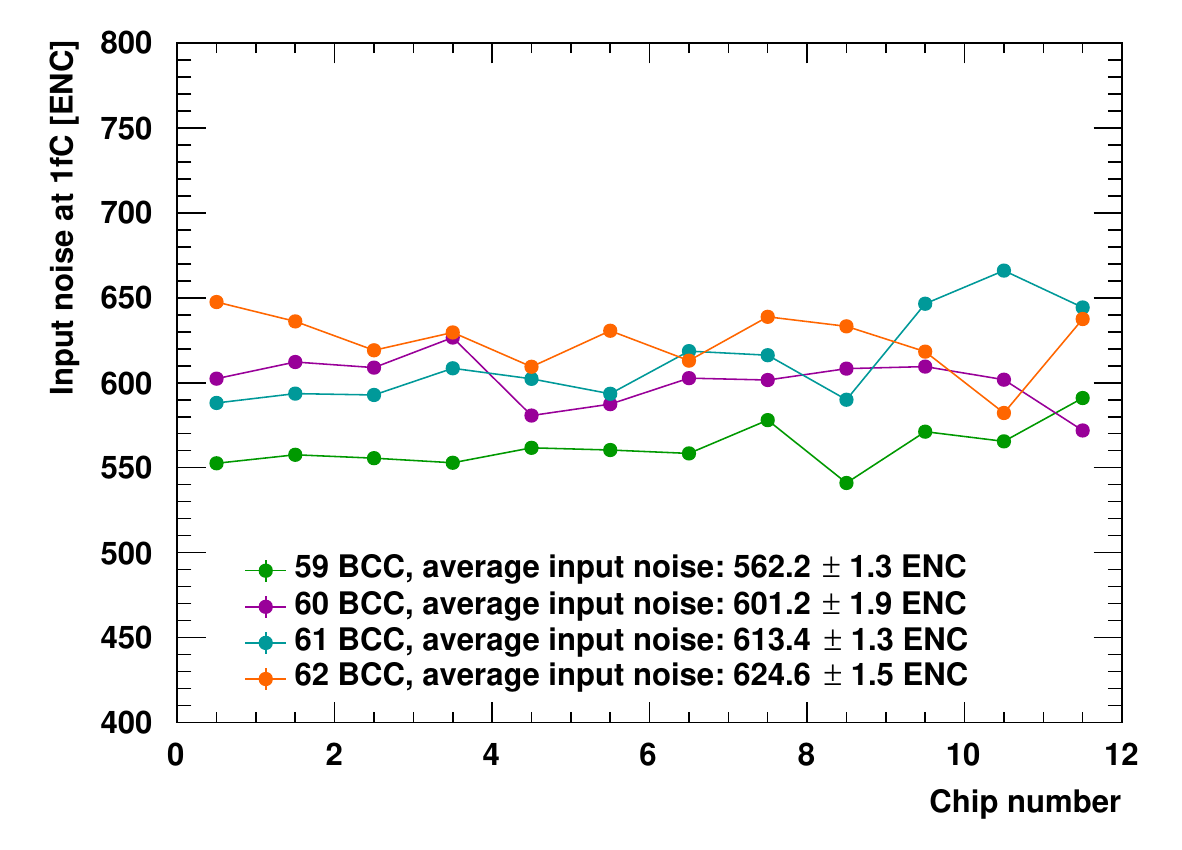}
    \caption{Averaged input noise over ten measurements for twelve ASICs for four modules of a \CommonRS\ \petalet\ read out simultaneously (\petalet\ common 1). The uncertainties are in the order 0.1\% and not visible in the histogram.} 
                \label{desypetalet}
\end{figure}

All modules of each \petalet\ are operational showing a successful assembly. The leakage currents are probed after assembly and have a similar behavior compared to the test of the modules on test frames.
A representative input noise figure obtained from four modules of one \petalet\ is presented in figure~\ref{desypetalet}. It shows the averaged input noise from ten measurements for twelve ASICs of each module at a bias voltage of 150\,V and cooling liquid temperature of 0$^{\circ}$C~\cite{nataliia}.
BCC adress 59, corresponds to the upper module on the front side, while BCC adress 60, to the lower one.
BCC adress 61 (upper module) and 62 (lower module) correspond to the modules of the back side.
Individual variations between modules are caused by different sensor qualities and different interstrip capacitances due to longer strip lengths for lower modules.
The uncertainties are in the order 0.1\% and not visible in the histogram.

\begin{figure}[h!]
    \centering
    \includegraphics[width=0.7\textwidth]{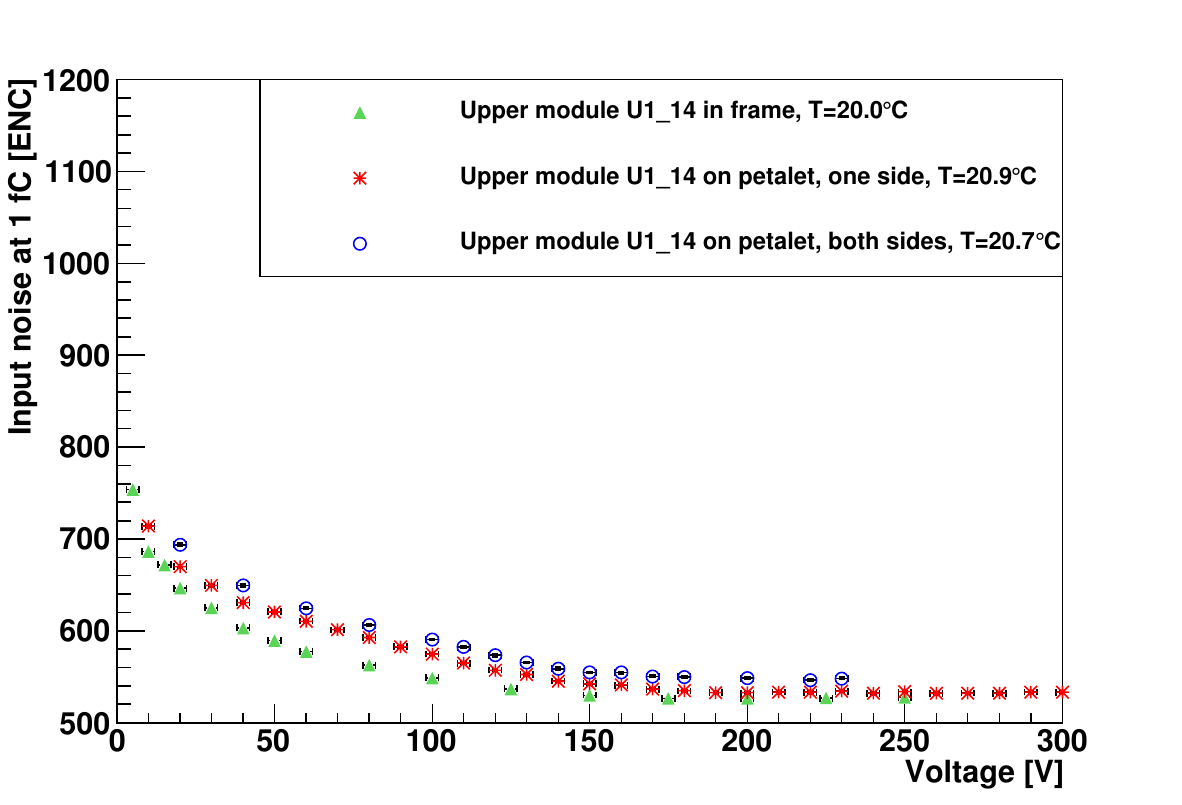}  
    \caption{Input noise distributions for a standard upper module on a \SplitRS\ \petalet\ as a function of bias voltage.
      The values are averaged over all chips on the hybrid of the module. Curves are given for operation in the test frame (green) and after assembly on the core, operating one side (red) and two sides (blue) (\petalet\ common 4).}
                \label{petalBT2}
\end{figure}

\begin{figure}[h!]
    \centering
    \includegraphics[width=0.7\textwidth]{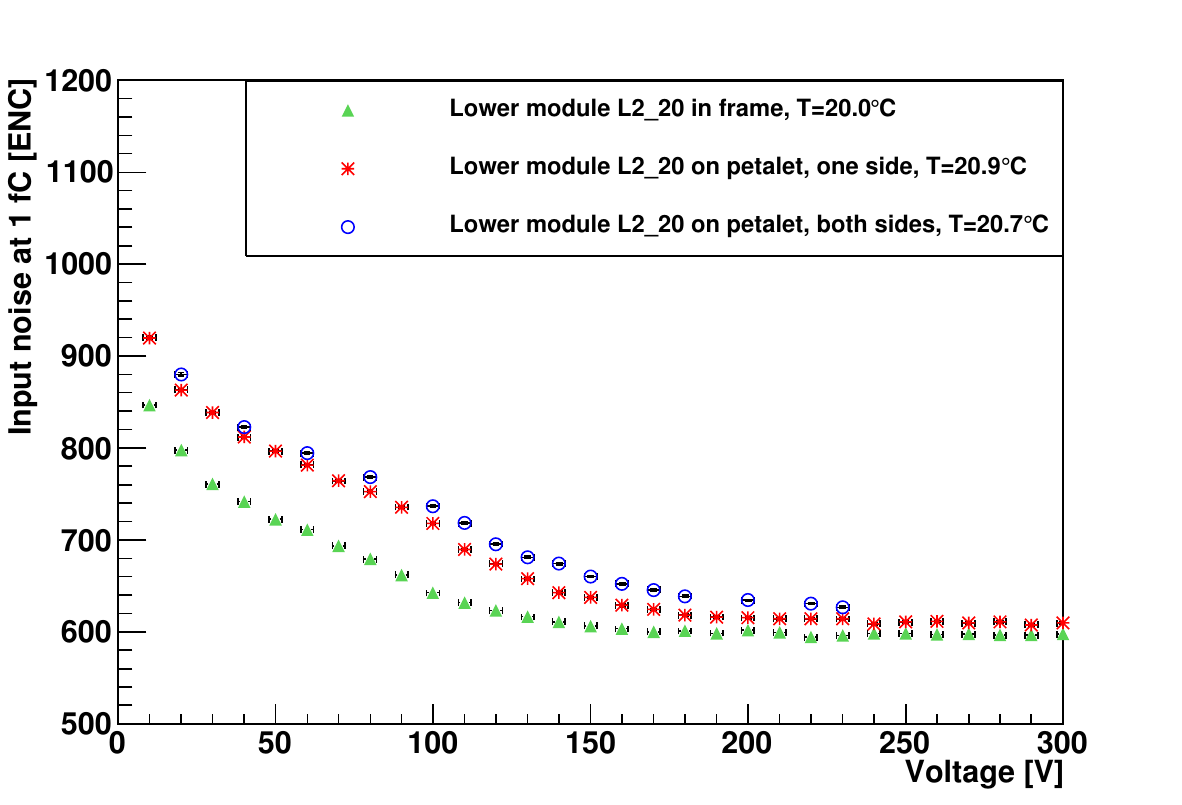}
    \caption{Input noise distributions for a standard lower module on a \SplitRS\ \petalet\ as a function of bias voltage.
      The values are averaged over all chips on the hybrid of one upper module. Curves are given for operation in the test frame (green) and after assembly on the core operating one side (red) and two sides (blue) (\petalet\ common 4).}
                \label{petalBTE}
\end{figure}

The comparison between the behavior of modules tested in test frames, on the core read out individually and simultaneously is given in figure~\ref{petalBT2} and~\ref{petalBTE}.
Figure~\ref{petalBT2} shows the input noise distribution for an upper standard module and figure~\ref{petalBTE} shows one for a lower module having embedded sensors.
Distributions of the average input noise on twelve ASICs are presented as a function of bias voltage. The green curves correspond to operation in the test frame, the red ones having one side of the \petalet\ operating and the blue ones having both sides powered and read out. 
The similarity of the red and blue data points shows that there is no cross-talk leading to changes in noise due to operation of one side or both sides of the \petalet.
In all modules the noise is slightly increased after assembly on the cores (red and blue curves) compared to test results in the test frame (green curves). 
After reaching full depletion, input noise values are about 550\,ENC for the upper and 650\,ENC for the lower module. 
Similarly full depletion of all sensors could be obtained for the other \petalets\ and the measurements show a stable operation by having variations in repeated runs of less than 5\,ENC. 

A stable operation of the modules is possible and the long-term behavior was tested over 12\,hours. Representative values for one \petalet\ are given in table~\ref{tablongtherm}.
\begin{table}[htp]
\begin{center}
\begin{tabular}{ccc}
\hline\hline
Module ID & Input Noise [ENC]  \\
\hline
upper module front & 763$\pm$2   \\
lower module front & 789$\pm$3   \\ 
upper module back & 647$\pm$1   \\
lower module back & 753$\pm$3   \\
\hline\hline
\end{tabular}
\end{center}
\caption{Average input noise values for all modules on a \petalet\ (common 5) in a long-term measurement of 12\,hours.
}
\label{tablongtherm}
\end{table}

The input noise values can be reduced by optimized sensors but also by improving the grounding and shielding. This has been pursued by both using shields on power lines of the bus tapes and around the coils of the DC-DC converters.

\subsubsection{Thermal tests}
\begin{figure}[h!]
    \centering
    \includegraphics[width=0.7\textwidth]{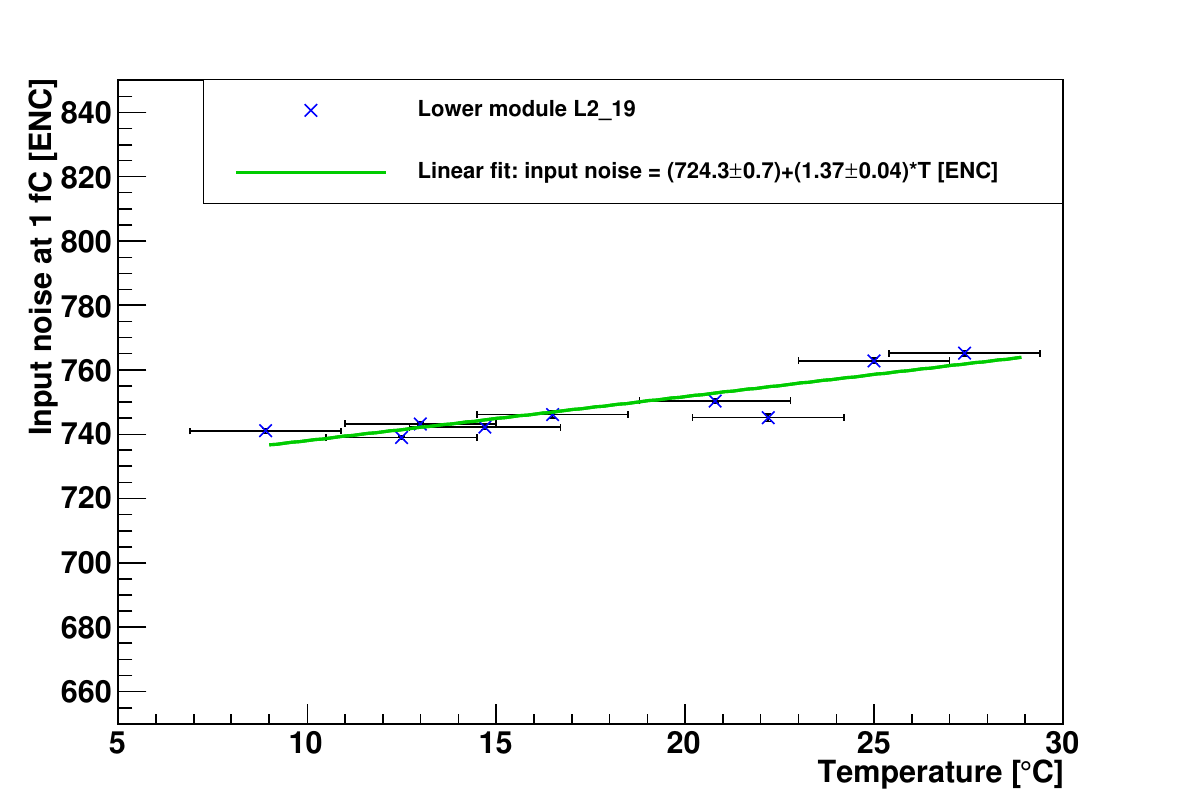}
    \caption{Averaged input noise of one module for different operation temperatures of the \petalet.}
                \label{petalBT1}
\end{figure}

Two types of thermal tests were performed. First, the temperature of the \petalet\ under operation was measured with NTCs at different positions of the core surface. 
The warmest area is found at the DC-DC converters.
Second, the input noise behavior was tested at different operation temperatures in an insulated box.
The result can be seen in figure~\ref{petalBT1} in the temperature range from 7$^{\circ}$C to 28$^{\circ}$C.
A slope of around 1.4\,ENC per degree is found deploying a linear fit to the data. 

\subsection{Testing of \CommonRS\ \petalet}
\label{subsec:testlf}
\begin{figure}[h!]
    \centering
    \includegraphics[width=0.7\textwidth]{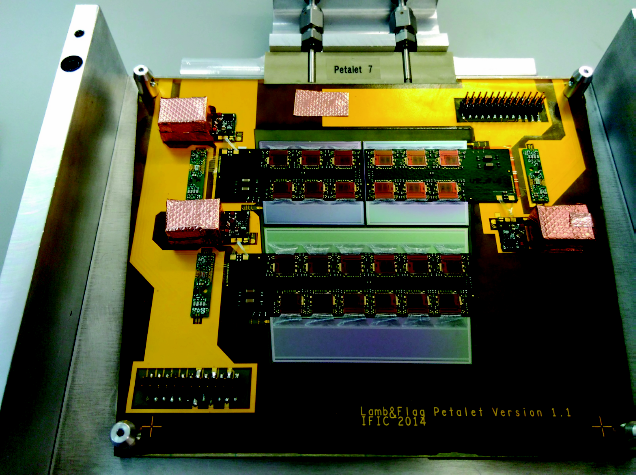}
    \caption{Photo of \CommonRS\ \petalet.}
                \label{lfpetal}
\end{figure}

\begin{figure}[h!]
    \centering
    \includegraphics[width=0.7\textwidth]{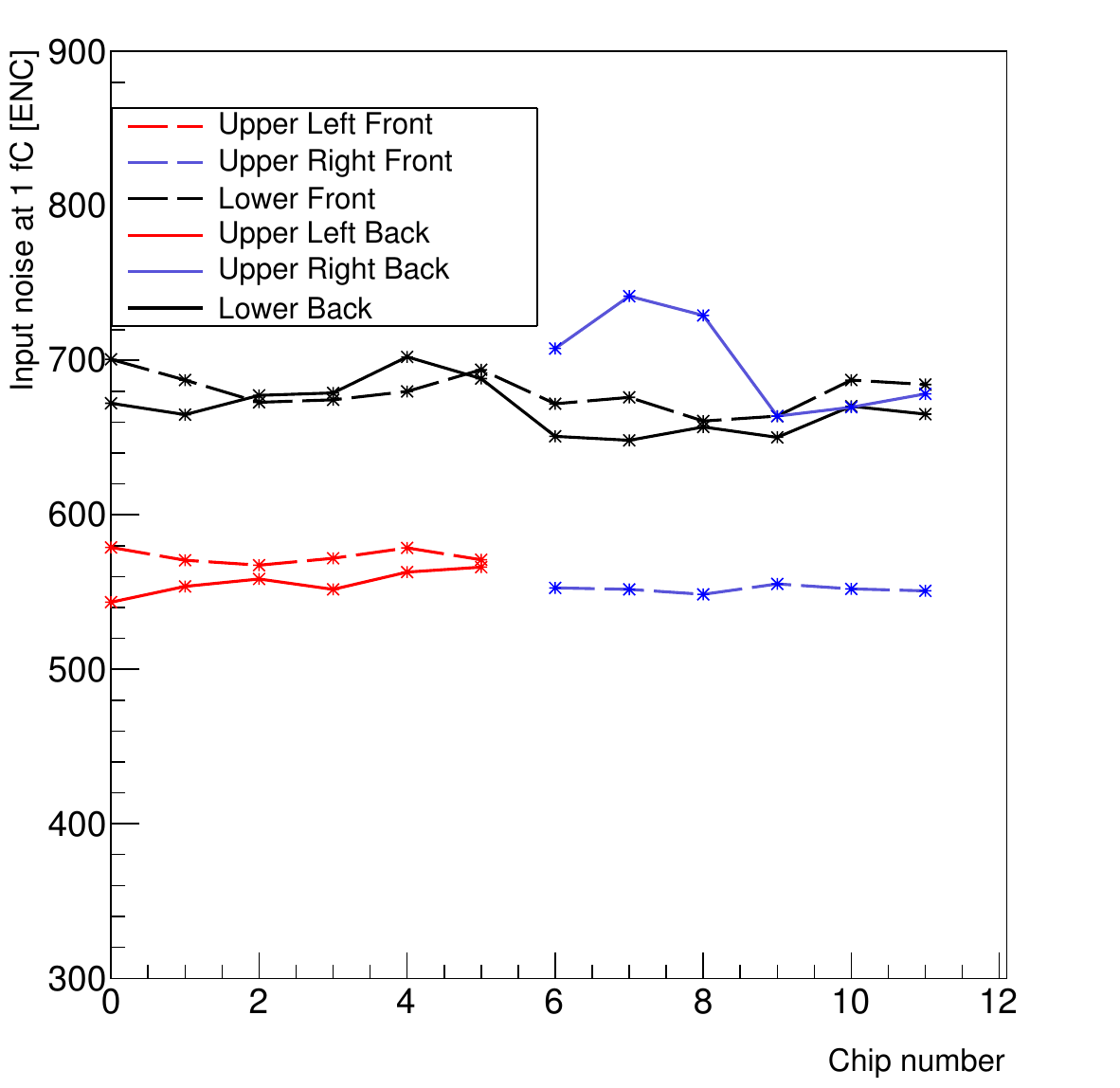}
    \caption{Input noise distributions for standard modules on \CommonRS\ \petalet\ for individual chips of all six modules at a bias voltage of 150\,V.
      The values are averaged over channels on each chip.} 
                \label{noiselfp}
\end{figure}

The \petalet\ in \CommonRS\ is shown in figure~\ref{lfpetal} and was evaluated with the same electrical tests. Three DC-DC converters and BCC boards, one for each module are placed next to each one on both core surfaces.
All six modules operated however not all BCC boards were fully functional and two were replaced with spares. The input noise distributions for individual chips of all six modules at a bias voltage of 150\,V which is above the full depletion voltage, are shown in figure~\ref{noiselfp}. The values are averaged over the 128 channels of each chip.
Variations between chips of one module are small. The upper right back module had problems during gluing which resulted in an increased noise compared to the other upper modules. 
A summary of the electrical performance is given in table~\ref{lfvalues}. 
\begin{table}[htp]
\begin{center}
\begin{tabular}{ccc}
\hline\hline
Module ID & Input noise [ENC]  \\
\hline
 upper left front & 572.0$\pm$0.7   \\
 upper right front & 559.4$\pm$0.6   \\ 
 lower front & 675.3$\pm$0.7   \\ 
 upper left back & 563.0$\pm$0.8   \\
 upper right back & 694$\pm$1.4   \\
 lower back & 665.1$\pm$0.7   \\ 
\hline\hline
\end{tabular}
\end{center}
\caption{Averaged input noise values of chips for six modules of the \CommonRS\ \petalet.
}
\label{lfvalues}
\end{table}

\subsection{Comparison of readout schemes}
\Petalets\ of both readout schemes were successfully assembled and tested.
Table~\ref{sumnoise} summarizes the electrical performance, including the averaged input noise values for each \petalet and indicating the operation conditions (sensor type and grade, maximum bias voltage, operation temperature). 
The measurements show that both schemes deliver adequate signals and are feasible to operate.
The input noise values vary between 590 and 730\,ENC. The variations are caused by different sensor grades.
In line with the results obtained with single modules, no large differences in the input noise performance are observed between the two readout schemes during multi-module tests.
\begin{table}[htp]
\begin{center}
\begin{tabular}{lcccccc}
\hline
\Petalet\ & common 1 & common 2 & common 3 & common 4 & common 5 & split 1  \\
\hline
sensor type  & std & std & emb & std & emb & std  \\
sensor grade & A & B & A & A/B & A/B & A  \\
operation  & &  & &  &  &  \\
temperature [$^{\circ}$C]  & 0& 15 & 5 & -20 & -20 & -15 \\
bias voltage [V]  & 150  & 150 & 140/130 & 200 & 100/250 & 150  \\
Input noise [ENC]  & 601 & 674 & 728 & 597 & 727 & 621  \\
\hline\hline
\end{tabular}
\end{center}
\caption{Overview of input noise values for all \petalets\ and operation conditions. Std. refers to standard sensors, emb to embedded sensors. Similarly in Ref.~\cite{nataliia}.
}
\label{sumnoise}
\end{table}
The behavior after irradiation was not the main scope of the \petalet\ programme and will be studied in detail once the final detector components are available (e.g. new ASICs).\newline
The \CommonRS\ has an electrical topology which covers both sides of a \petalet. This implies that the bus tape is folded on top of the core but only one end-of-petalet board is required. 
The later is also challenging giving electrical routing constraints. 
It has the advantage of only one bus tape flavor and allows a reduced amount of tape since not all area of the \petalet\ requires coverage. 
The bus tapes of the \SplitRS\ are single-sided bus tapes and required in two different types to connect the modules on the front and back sides to separate end-of-structure boards. Whilst this concept results in a larger tape surface, it requires a smaller amount of DC-DC converters and offers advantages in terms of HV insulation and ease of routing.\newline
A comparison of the mechanical implications of the two readout schemes shows that  
for assembly and particularly the gluing procedure, the bus tape of the \CommonRS\ was slightly more difficult due to its wrapping around the top of the carbon core. In addition, its long shape could enhance CTE mismatches. Similar difficulties are expected when co-curing bus tapes on face sheets. 
The \SplitRS\ bus tapes have slightly more material because they have also kapton below the sensor area which can be avoided in future for bus tapes in \CommonRS. It is estimated to be an increase of 3.5\% of the bus tape radiation length per bus tape for full-size petals.
In terms of module loading, no differences are observed between both readout schemes, as an equal number of mounting steps per module is required. Moreover, the assembly of the \SplitRS\ hybrids is easier in the module assembly step. 
For both schemes a higher accuracy in module placement is expected making use of the sensor fidiucials.\newline
Modules are single entities in both concepts and it is suggested to reduce the width of the hybrids to match the sensor widths~\cite{modulepaper}.

In summary, despite slightly additional material and the need of different bus tape flavours per side, after an ITk internal review, the ITk strips community agreed to use the \SplitRS\ for future prototyping of large structures, i.e. petals.  
This choice will also allow more direct synergies with the developments of the central part of the future ATLAS strip tracker.

\section{Summary and conclusions}
\label{sec:result}
In the \petalet\ project, fully equipped structures for the upgrade of the forward region of the silicon strip tracker of the ATLAS experiment were developed, prototyped and evaluated.
In addition, the project aimed for the demonstration of the feasibility of the components and a selection of the readout scheme.
Six fully integrated \petalets\ were successfully built deploying two different readout schemes.
The feasiblity of which has been demonstrated.\newline
Carbon-fibre sandwich structures with embedded titanium cooling pipes were built, including the development of bent pipes. Two bus tape designs were laid out and the gluing process of the bus tapes onto the cores was developed. 
In the future, co-curing of the bus tapes and carbon fibre facesheets is expected to improve the planarity of the assembly.
Preliminary prototyping of this process shows very encouraging results.
The silicon sensor modules were mounted on the cores by custom-designed tooling, maintaining a precision to keep clearances between modules.   
The introduction of fiducials in both the silicon sensors and the petal cores is expected to enhance the positioning accuracy of the modules.
The thermal behavior of the \petalets\ is acceptable as expected in the measured range.
The modules of the various \petalets\ have been found to be fully functional after loading and the development of adpater boards was succesful. Both sides of the \petalets\ can be read out simultanously and do not influence each other largely. The direct comparison of input noise values of different \petalets\ is misleading due to different sensor grades which dominate the noise behavior. In summary, the mean input noise values averaged over all modules of one \petalet\ show variations for different \petalets\ between 590\,ENC and 730\,ENC. The minimum input noise of a lower module is 650\,ENC and for an upper one 550\,ENC in both layouts.
Comparing these values to results from single modules measured in test frames (lower modules about 650\,ENC, upper modules about 520\,ENC)~\cite{modulepaper}, the noise values are slightly higher for \petalets.
The grounding was optimized for all \petalets.
General problems in the operation were the initial start of modules and the operation of BCC boards which partially failed and needed to be replaced. For modules on petals they will be replaced with new ASICs and implemented in the hybrids. 
The connectors soldered on the bus tapes proved to be fragile, and thus their use is envisioned to be avoided in future bus tapes.

The experience and knowledge gained is directly transferred to the prototyping of full-size petals. They will employ the \SplitRS\ due to easier module assembly and bus tape routing.
In future studies irradiation, systematic temperature cycling and high rate tests are foreseen.

\section*{Acknowledgements}
This work is supported and financed by the German Federal Ministry of Education and Research, the Helmholtz Association, Germany, the Spanish Ministry of Economy and Competitiveness through the Particle Physics National Program (ref. FPA2012-39055-C02-01, FPA2012-39055-C02-02, FPA2015-65652-C4-1-R and FPA2015-65652-C4-4-R) and co-financed with FEDER funds, the European Social Fund, by the Ministry Of Science, Research and Arts Baden-Wuerttemberg, Germany, and the UK Science and Technology Facilities Council (under Grant ST/M006409/1).



\begin{thebibliography}{19}
\bibitem{stripTDR} The ATLAS Collaboration, \emph{Technical Design Report for the ATLAS Inner Tracker Strip Detector}, CERN-LHCC-2017-005, ATLAS-TDR-025 (2017).
\bibitem{sergio} S.~Diez, et al, A double-sided, shield-less stave prototype for the ATLAS Upgrade strip tracker for the High Luminosity LHC, \emph{JINST} {\bf 9} P03012 (2014).  
\bibitem{sensorpaper} V.~Benitez et al., \emph{Sensors for the End-Cap Prototype of the Inner Tracker in the ATLAS Detector Upgrade}, \emph{Nucl. Instrum. and Meth. A}  {\bf 833} (2016) 226. 
 \bibitem{modulepaper} S.~Kuehn, et al., Prototyping of hybrids and modules for the forward silicon strip tracking detector for the ATLAS Phase-II upgrade, \emph{JINST} {\bf 12} P05015 (2017). 
\bibitem{sven} M.~Aliev, et al.,A forward silicon strip system for the ATLAS HL-LHC upgrade, Nucl. Instrum. and Meth. A730 (2013) 210--214.
\bibitem{kambiz} K.~Mahboubi et al., \emph{The front-end hybrid for the ATLAS HL-LHC silicon strip tracker}, \emph{JINST} {\bf 7} C02027 (2014).
\bibitem{kaplon} J.~Kaplon, \emph{The ABCN Front-end Chip for ATLAS Inner Detector Upgrade}, Topical Workshop on Electronics for Particle Physics (TWEPP 2008), pg. 116, [http://cdsweb.cern.ch/record/1158514].
\bibitem{abcn} W.~Dabrowsky et al., \emph{Design and performance of the ABCN-25 readout chip for the ATLAS inner detector upgrade}, \emph{IEEE Nucl. Sci. Symp. Conf. Rec.} {\bf 2009} (2009) 373.
\bibitem{abc130} N.~Lehmann, \emph{Tracking with self-seeded Trigger for High Luminosity LHC}, Master's thesis, 2014, [https://documents.epfl.ch/users/n/nl/nlehmann/www/SelfSeededTrigger\_MasterThesis/SelfSeeded\newline Trigger\_NiklausLehmann\_Thesis.pdf].
 \bibitem{CNM} IMB-CNM, Instituto de Microelectronica de Barcelona, Centro Nacional de Microelectronica, 08193 Cerdanyola del Valles (Bellaterra), Barcelona, Spain.
\bibitem{embedded_upgrade} M.~Ull\'{a}n et al., \emph{Embedded pitch adapters for the ATLAS Tracker Upgrade}, \emph{Nucl. Instrum. and Meth. A} {\bf 732} (2013) 178.
\bibitem{hiroemb} M.~Ull\'{a}n et al., \emph{Embedded Pitch Adapters: a High-Yield Interconnection Solution for Strip Sensors}, \emph{Nucl. Instrum. and Meth. A} {\bf 831} (2016) 221.
\bibitem{allcomp} Allcomp Inc.  \emph{http://www.allcomp.net/}
\bibitem{henkel} Henkel Adhesives International, [http://www.henkel-adhesives.com/henkel-adhesives.htm].
\bibitem{dupont} DuPont Inc.  \emph{http://www.dupont.com/}.
\bibitem{elgo} Elgoline d.o.o. \emph{http://www.elgoline.si/}.
\bibitem{IPC} IPC Association Connecting Electronics Industries, \emph{IPC-2223A: Sectional Design Standard for Flexible Printed Boards}, June 2004. 
 \bibitem{hsio} D.~Nelson, \emph{HSIO Development Platform Users Guide Version C02}, Revision 1.1 Edition, April 2010, \emph{http://www.slac.stanford.edu/djn/Atlas/hsio}.
 \bibitem{dowcorning} Dow Corning Inc. \emph{https://www.dowcorning.com/}.
 \bibitem{nataliia} N.~Zakharchuk, \emph{Measurement of $Z$-boson production cross sections at $\sqrt{s}=$13\,TeV and $t\bar{t}$ to $Z$-boson cross-section ratios with the ATLAS detector at the LHC}, PhD thesis, 2017.
\end{thebibliography}
\end{document}